\documentclass[usenatbib]{mnras}
\usepackage{epsfig}
\usepackage{amsmath,amssymb}
\usepackage{xcolor}
\usepackage{booktabs}
\usepackage{pdflscape}
\usepackage{textcomp}

\usepackage{soul}
\usepackage{aas_macros}  
\usepackage{chngcntr}
\graphicspath{{figures/}}

\newcommand{\Msolar}{\mbox{\,$\rm M_{\odot}$}}        

  \newcommand{\Teff}{\mbox{\,\em T$_{\rm eff}$}}         
  \newcommand{\sg}{\mbox{\,log $g$}}                     
 \newcommand{\teff}{\mbox{\,$T_{\rm eff}$}}      
\newcommand{\lgcs}{\mbox{\,$\log g / {\rm cm\,s^{-2}}$}}        

  \newcommand{\iso}[2]{\mbox{$^{#1}{\rm #2}$}}           
  \newcommand{\vsini}{\mbox{\,$v\,\sin i$}}              
  \newcommand{\kmsec}{\,\mbox{$\mbox{km}\,\mbox{s}^{-1}$}}    
  \newcommand{\msec}{\,\mbox{$\mbox{m}\,\mbox{s}^{-1}$}}    
 \newcommand{\cmss}{\,\mbox{$\mbox{cm}\,\mbox{s}^{-2}$}}    
  \def\simge{\mathrel{\raise1.16pt\hbox{$>$}\kern-7.0pt
    \lower3.06pt\hbox{{$\scriptstyle \sim$}}}}           
  \def\simle{\mathrel{\raise1.16pt\hbox{$<$}\kern-7.0pt
    \lower3.06pt\hbox{{$\scriptstyle \sim$}}}}           

\title[DY Cen revisited]{SALT revisits DY\,Cen: a rapidly-evolving strontium-rich single helium star}

\author[C. S.~Jeffery et al.]{C. Simon Jeffery$^{1,2}$\thanks{email: simon.jeffery@armagh.ac.uk}, N. Kameswara Rao$^3$, and David L. Lambert$^4$ \\
$^{1}$Armagh Observatory and Planetarium, College Hill, Armagh BT61 9DG, UK\\
$^{2}$School of Physics, Trinity College Dublin, College Green, Dublin 2, Ireland\\
$^{3}$Indian Institute of Astrophysics, Bangalore 560034, India\\
${^4}$The W. J. McDonald Observatory and Department of Astronomy, University of Texas at Austin, Austin, TX 78712-1083, USA
}
\begin{document}

\date{Accepted \ldots. Received \ldots; in original form \ldots}

\pagerange{\pageref{firstpage}--\pageref{lastpage}} \pubyear{2014}

\maketitle

\label{firstpage}

\begin{abstract}
The hydrogen-deficient star DY\,Cen has been reported as an R\,CrB-type variable, an extreme helium star (with some hydrogen), and as a single-lined spectroscopic binary. 
It has been  associated with a dramatic change in visual brightness and colour corresponding to a change in effective temperature (\Teff) of some 20\,000\,K in the last century. 
To characterize the binary orbit and \Teff\ changes more precisely, new high-resolution spectroscopy has been obtained with SALT. 
The previous orbital period is not confirmed; previous measurements may have been confused by the presence of pulsations. 
Including data from earlier epochs (1987, 2002, and 2010), self-consistent spectral analyses from all four epochs demonstrate an increase in \Teff\ from 18\,800 to 24\,400\,K between 1987 and 2015. Line profiles demonstrate that the surface rotation has increased by a factor two over the same interval. This is commensurate with the change in \Teff\ and an overall contraction. Rotation will exceed critical if contraction continues. 
The 1987 spectrum shows evidence of a very high abundance of the s-process element strontium. 
The very rapid evolution, non-negligible surface hydrogen and high surface strontium point to a history involving a very late thermal pulse. Observations over the next thirty years should look for a decreasing pulsation period, reactivation of R\,CrB-type activity as the star seeks to shed angular momentum and increasing illumination by emission lines from nebular material ejected in the past.    
\end{abstract}


\section{Introduction}
\label{s:intro}
DY\,Cen has been identified as an extreme helium star having above average surface hydrogen abundance \citep{jeffery93a}.
The star showed R\,CrB-type variations up until 1934, but not since \citep{hoffleit30,schaefer16}. 
Its evolution appears to be rapid: a steady decline in visual  brightness over about the last century \citep{demarco02,schaefer16} and an increase of effective temperature of about  300 K yr$^{-1}$ in the last 60 years \citep{pandey14,schaefer16} are two indicators of evolution. The recent claim by \citet{rao12} based on a compilation of radial velocities over several decades that DY\,Cen is a 39.7\,d eccentric single-lined  spectroscopic binary with a stellar separation of about 10/$\sin i R_\odot$ is not readily reconciled with the star’s rapid evolution.  Since stars such as extreme helium and R\,CrB-like stars are expected to evolve rapidly at approximately constant luminosity, the increase in effective temperature implies a decrease in stellar radius with the implication that the binary components should have been in contact only about 100 years ago or less and, perhaps, a common-envelope binary was then present.  These startling conclusions led to our acquisition of a sequence of  high-resolution spectra with the Southern Africa Large Telescope (SALT)  presented in this paper. Not only do we provide a fresh examination of the radial velocity evidence for a spectroscopic binary but we rediscuss the stellar effective temperature, surface gravity and atmospheric composition of this rather special star.

\section{Observations}
\label{s:obs}
Observations obtained with the ultraviolet \'echelle spectrograph (UVES: $R\approx40\,000$) on the European Southern Observatory (ESO) Very Large Telescope (VLT) and previously reported by \citet{rao12} were re-extracted in fully reduced format (i.e. with orders stitched) from the ESO archive.  

New observations were obtained with
the High Resolution Spectrograph \citep[HRS: $R\approx43\,000$, $\lambda\lambda = 4100 - 5200$\AA][]{bramall10} of SALT during 2015 and 2016 on the dates shown in Table 1. 
The spectra were reduced to order-by-order wavelength calibrated rectified form using the SALT pipeline pyHRS \citep{crawford16}; orders were stitched into a single spectrum using our own software. In general, spectra were obtained in pairs (or a higher multiple) 
which were coadded to provide a single observation for each date.

\begin{table}
\begin{center}
\caption{Measurements of the DY\,Cen radial velocity. New measurements are labelled `2020', values reported by \citet{rao12} are labelled `2012'. Errors in the last significant figure(s) are given in parentheses. For the 2020 measurements, these are errors in the relative velocities. Systematic errors in the absolute velocities are discussed in the text.}  
\label{t:vels}
\begin{tabular}{clll}
\hline
 Instrument & HJD & $v$ \kmsec &  \\
  & -- 2400000 &  2020  &  2012  \\
  \hline
CTIO 4m & 45068.734 & & 35(3) \\
CTIO 4m & 45070.624 & & 25(3) \\
CTIO 4m & 45070.644 & & 25(3) \\ 
CTIO 4m & 45071.694 & & 32 3) \\
AAT/RGO & 47233.774 & & 15.1(2.5) \\ 
CTIO 4m & 47723.70 & & 41(4) \\
CTIO 4m &  48762.608 & & 29(4) \\ 
AAT/UCLES & 51215.248 & & 37.3(2.4) \\ 
AAT/RGO & 52003.998 & & 31.7(2.7) \\
AAT/RGO & 52005.137 & & 32.5(2.6) \\ 
AAT/UCLES & 52449.926 & & 23.8(2.0) \\ 
AAT/UCLES & 52838.903 & & 40.0(1.5) \\[1mm] 
VLT/UVES & 55254.809 & 22.91(15) & 24.7(2.6) \\ 
(2010) & 55257.758 & 23.37(76) & 25.4(2.5) \\
     & 55260.803 & 32.57(06) & 36.9(4.6) \\
     & 55280.608 & 17.21(55) & 14.2(3.6) \\[1mm]
SALT/HRS & 57141.309  & 35.15(36)   &  \\
(2015/6) 
    & 57143.270  &  32.70(29)  &  \\
    & 57147.258  &  28.07(31)  &  \\
    & 57159.238  &  34.61(26)  &  \\
    & 57186.398  &  33.41(29)  &  \\
    & 57195.395  &  24.67(28)  &  \\
    & 57205.371  &  36.23(29)  &  \\
    & 57214.313  &  22.49(32)  &  \\
    & 57232.266  &  31.45(25)  &  \\
    & 57535.465  &  16.30(29)  &  \\
    & 57564.348  &  26.56(29)  &  \\
\hline
\end{tabular}
\end{center}
\end{table}

\begin{figure*}
\epsfig{file=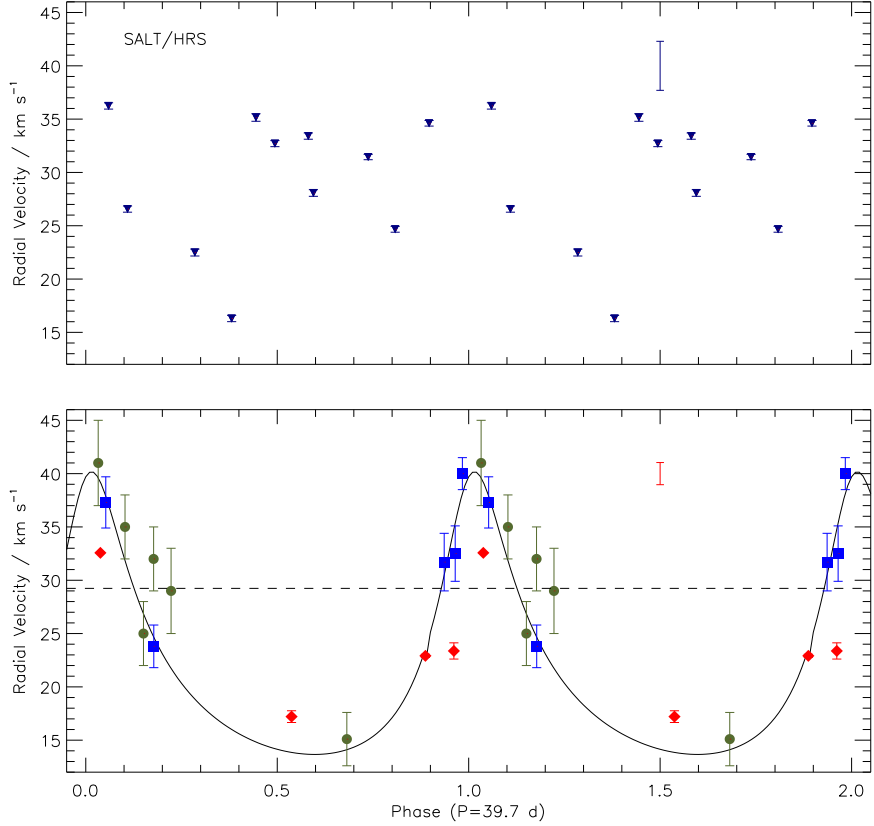,width=120mm,clip=,angle=0}
\caption{Radial velocities of DY\,Cen phased on the ephemeris of \citet{rao12} and folded over two cycles. Velocities measured from HRS (inverted triangles) are shown in the top panel. The lower panel shows revised measurements from  UVES (red diamonds), Herbig (green circles) and the AAT (blue squares) and the orbital solution (smooth curve) from \citet{rao12}. The mean value of the HRS velocities is shown as a dashed line. The systematic or template  errors on the HRS (top: blue) and UVES (bottom: red) velocities are shown without symbols.} 
\label{f:orbit}
\end{figure*}

\section{Radial velocities}
\label{s:vels}
\subsection{Measurement}
Radial velocities were obtained from SALT/HRS spectra by the method of cross-correlation.

It is useful first to reprise the procedure used to construct the cross-correlation function. Both spectra to be compared should be rectified to the continuum, which must then be subtracted so that the line spectrum represents a signal with a reference value of 0. Each spectrum should be binned such that each increment represents an equal step in the logarithm of the wavelength ($\ln \lambda$)\footnote{Our linearisation of our SALT/HRS spectra converts directly from instrument pixels in the dispersion direction to equal increments in $\ln \lambda$, thus avoiding an extra rebinning step prior to cross-correlation.}. This is because velocity shift is a logarithmic function of wavelength ($\delta v / c = \delta \lambda / \lambda = \delta \ln \lambda $). The cross-correlation function (ccf) for discrete signals such as digitized astronomical spectra is defined 
\[
(f\times g)[n] = \sum_{m=-\infty}^{\infty} \overline{f(m-n)}g(m).
\]
The ccf of two similar signals separated by a uniform shift normally appears as a near-symmetric peak with noise either side. The position of the peak maximum is obtained by optimizing a parabola to points around the peak maximum and represents the velocity shift $\delta v$ (in appropriate units). A formal error estimate is provided by the errors in the parabola fit coefficients.   
The power of the method for measuring  radial velocities from astronomical spectra lies in combining information from all spectral lines in the wavelength range observed and hence gives excellent results even when low signal-to-noise in individual spectra precludes the direct measurement of line shifts.

\begin{figure*}
\epsfig{file=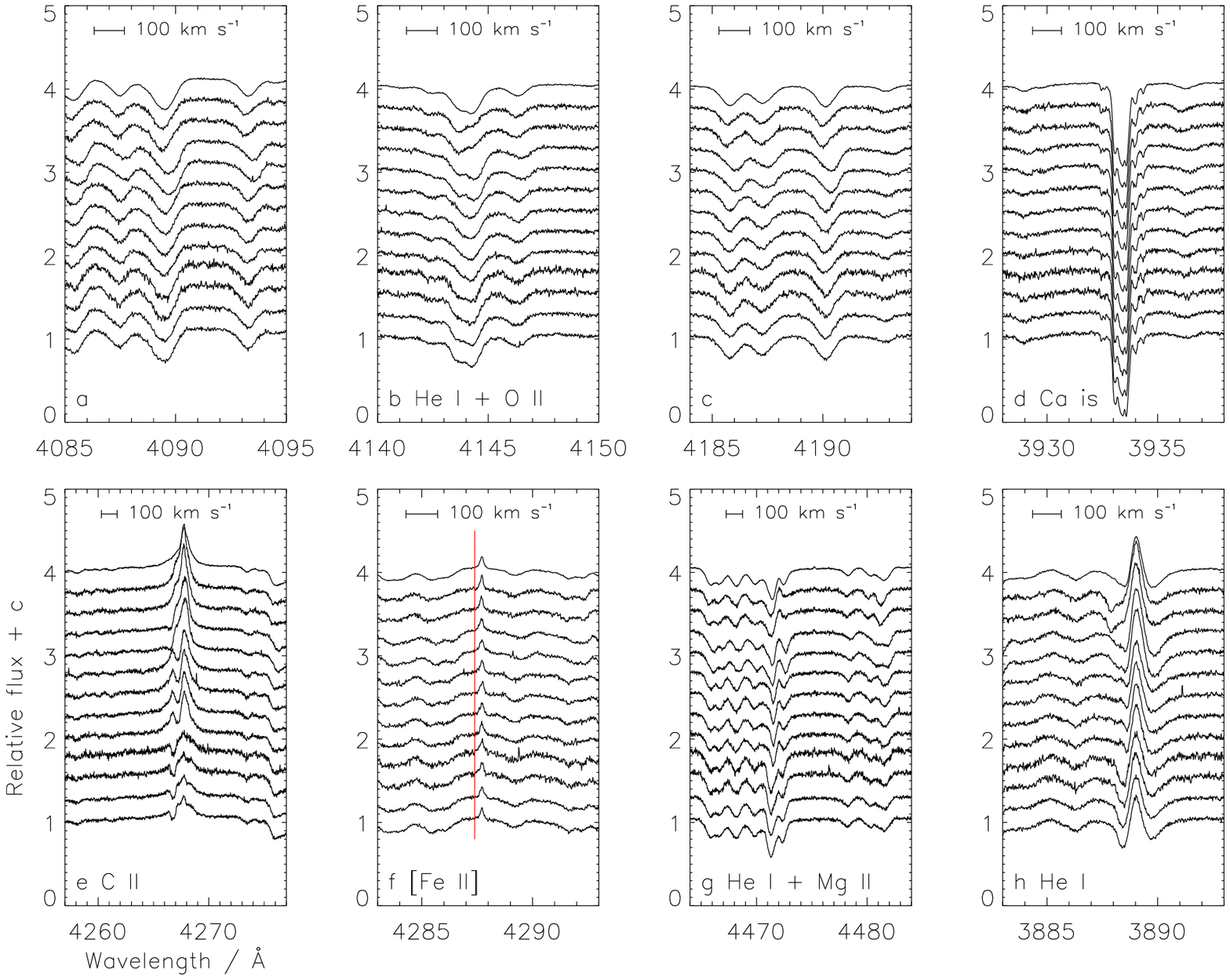,width=175mm,clip=,angle=0}
\caption{Time-series VLT/UVES spectroscopy of DY\,Cen in selected spectral windows showing (a) - (c) photospheric line-profile variations commensurate with a radial pulsation; (b) includes a blend of He{\sc i} 4143\AA\ with several O{\sc ii} lines, (d) a stationary multi-component interstellar absorption line, (e) variable circumstellar emission (C{\sc ii}), (f) a stationary [Fe{\sc ii}] nebula line (rest wavelength marked in red), (g) a complex region including photospheric Mg{\sc ii} and photospheric and circumstellar He{\sc i}, and (h) strong stationary circumstellar He{\sc i} emission overlying variable photospheric absorption. All spectra have been corrected for earth motion and are normalised to unity. They are offset vertically in increments of 0.25, with time running upwards. The mean spectrum is shown at the top. 
Mid-times of observations were (BJD-2455200) 54.770, 54.792, 54.826, 54.848,57.719, 57.741, 57.775, 57.796, 60.793, 60.814, 80.598 and 80.619. The bar at the top of each panel represents a velocity shift of $100 \kmsec$. }
\label{f:lpvs}
\end{figure*}

In the present case, measurements from individual observations of DY\,Cen were obtained by cross-correlation with a reference spectrum (or template) that was itself a high signal-to-noise spectrum of DY\,Cen, being the sum of all those observed with the same instrument and configuration \citep[cf. ][]{jeffery86}. 
Such a `self-template' is constructed iteratively. 
In the first iteration, the template is a simple sum weighted by the mean photon counts recorded in each observation; any relative shifts in the constituent spectra are included and lines in the template are consequently broadened relative to lines in the constituent spectra. 
The first radial velocities are obtained by cross-correlation against this template. 
The second iteration starts by using these first velocity estimates to apply a shift to each constituent spectrum before addition, resulting in a new template that more closely resembles the stellar spectrum unbroadened by velocity shifts\footnote{Radial velocity shifts may be due to the acceleration of the star or of the observer or both.}. Repeating cross-correlation against this new template results in more precise measurements with smaller errors. 
The process can be repeated until measurements against successive templates have converged. 
In practice, only two iterations are found to be necessary, the third invariably producing results indistinguishable from the second.  
Since this method only provides results relative to the adopted template, 
the radial velocity of the latter must be obtained by cross-correlation with the spectrum of either a standard star of known velocity or a theoretical spectrum at laboratory rest velocity or by some other means. If either of the former, the reference spectrum should be chosen to resemble the target spectrum as far as possible so as to maximise the height and minimise the width of the ccf peak.  
For DY\,Cen, we used theoretical spectra similar to those obtained from the fine analysis described below. 

Best results are usually obtained from large regions of spectrum containing many sharp narrow lines. 
These are not always available in the case of early type stars where Stark broadening and variable emission render hydrogen and helium lines as sources of systematic noise rather than signal, and where thermal broadening and rotation contribute substantially to the widths of the remaining lines.  
Therefore the SALT/HRS velocities for DY\,Cen were obtained by cross-correlation using spectral segments of different sizes and wavelength ranges. 
Shifting the range by  $\pm 200 - 400$\AA\ had negligible effect  on the relative velocities; reducing the range by a factor 2 or more increased both scatter and formal errors. 
The SALT/HRS measurements were finally adopted from cross-correlation over the wavelength range 4400--5200\AA,  which includes more than 230 lines with equivalent width $W_{\lambda}>10$m\AA. Hydrogen and carbon emission lines were excluded since these do not share the motion of the photosphere.
With the HRS spectra (over)sampled at $R\approx192000$, 9 points in the ccf peak were used to determine the position of ccf maximum. 
9 points correspond to a window 14\kmsec\ wide; the typical full-width half maximum for strong absorption lines ({\it e.g.} Si{\sc iii}) in the HRS spectrum is 1.1\AA, or 72\kmsec. 
Extensive experiments with the sampling rate and the width of the window consistently give differences  $<1\kmsec$.
The template velocities were measured by cross-correlation with a theoretical spectrum at rest velocity.
Here, the choice of spectral range led to differences of up to 5\kmsec, suggesting residual errors in the overall wavelength calibration in some parts of the spectrum. 
This is treated as a systematic error $\approx\pm2.3\kmsec$ to be applied to all measurements. 
All velocities were corrected to the Solar System barycenter, and are reported in Table \ref{t:vels}. 

The HRS velocities are robust to the data reduction process. 
Reduced spectra from two different pipelines for the same data obtained on JD 2457535  give radial velocities different by $<40 \msec$. 
Since 2016, HRS radial velocities have been better than $300\msec$ for data products reduced with the {\sc midas} pipeline \citep{Kniazev16,kniazev16.hrs-rv}.
HRS calibrations obtained before 2016 were not always obtained consistently (Kniazev, private communication), and data products from the {\sc pyraf} pipeline \citep{crawford16} have shown sporadic velocity errors \citep{jeffery19b}. 
The HRS velocities for DY\,Cen are however internally consistent, as demonstrated by measurements of the [Fe{\sc ii}] 4287\AA\ line discussed in the next section. 

\subsection{VLT/UVES}
Previous measurements from VLT/UVES spectra were obtained using line-by-line measurements from individual \'echelle orders \citep{rao12}.
As an independent check, the cross-correlation method was used to obtain radial velocities from the re-extracted spectra, as for the SALT data. 
The VLT/UVES spectra were obtained on four separate nights, with multiple exposures taken on each night as reported by \citet{rao12}. Radial velocities were obtained for each individual blue spectrum.
 17 points in the ccf peak were used to determine the velocity and the associated formal error.
With an (over)sampling rate corresponding to $R\approx280\,000$, this corresponds to a window some 18 \kmsec\ wide around the cores of lines having a typical full-width half maximum of 0.9\AA\ or 58\kmsec. 
The systematic error in the template velocity is $\approx\pm1.0$\kmsec. 
Repeat observations  were combined to show the mean velocity for each night, together with the standard error  ($\sigma/\sqrt(n-1)$). The results tabulated in Table \ref{t:vels}, were obtained for the spectral range $4000 - 4150$\AA\  containing some 136 lines with $W_{\lambda}>10$m\AA,  together with the values reported by \citet{rao12}. 
Both sets of measurements are consistent with one another.
A cross-check was carried out using the cross-correlation method on two spectral windows restricted entirely to the interstellar calcium H and K lines. 
The standard deviations of all 12 measurements for each line were 0.63 and 0.58\,\kmsec\ respectively, with corresponding maximum ranges of 1.68 and 1.50\,\kmsec.

The [Fe{\sc ii}] 4287 and 4359 \AA\ emission lines in the VLT/UVES and SALT/HRS spectra are also stationary. 
The radial velocity of [Fe{\sc ii}] 4287.393 \AA\ was measured by fitting a parabola to the line peak to give means and standard deviations $22.55\pm0.80$ and $21.93\pm3.33$ \kmsec\ for the two datasets from 12 and 9 measurements respectively. 
The difference of $0.62$ \kmsec\ is well within the observational scatter, implying no detectable zero-point difference between VLT/UVES and SALT/HRS measurements.
The small scatter of Ca{\sc ii} and [Fe{\sc ii}] velocities supports the statement that nebular lines in DY\,Cen spectra show no radial velocity variation \citep{rao12}. 

Other radial velocities obtained prior to  2010 and reported by \citet{rao12} are also reproduced in Table \ref{t:vels}.
Their unweighted mean is $29.3\pm8.0$ \kmsec. The mean radial velocity from 2015/16 is $31.9\pm1.6$ \kmsec. 


\subsection{Analysis}

 The SALT/HRS velocities vary over a range of some 20\kmsec. 
 They show no evidence of behaviour consistent with the spectroscopic binary ephemeris published by \citet{rao12} drawing on radial velocities from 1982 to 2010. 
This inconsistency is well displayed in Fig.\,\ref{f:orbit} where velocities are placed on the ephemeris  from \citet{rao12} and the upper panel shows the HRS velocities and the lower panel shows the velocities assembled by Kameswara et al.  but with their UVES velocities replaced by those from Table\,\ref{t:vels}.  
That the 2012 collection of velocities reflects  real changes in mean velocity of absorption line profiles appears certain given that the velocity of the nebular emission lines is constant to within $\pm1.5\kmsec$ from 1992 to 2010 and furthermore their mean velocity is unchanged for the HRS spectra; the nebular emission lines originate from a low-density region likely remote from DY\,Cen and unaffected by atmospheric changes \citep{giridhar96}.

Resolution of the conflict between the velocity behaviour in the upper and lower panels of Fig.\,\ref{f:orbit} would seem to suggest two limiting situations: 
(i) DY Cen is not a single-lined spectroscopic binary and velocity variations reflect irregular pulsations on a timescale of several days; 
(ii) the 2012 paper did indeed show that DY\,Cen was a spectroscopic binary  but either atmospheric changes in 2015-2016 mask the binary signature (unlikely!) or the low-mass companion to DY Cen has been stripped of mass in a few years (also, unlikely!). 
Considering that the observed contraction rate and proposed orbit of DY\,Cen imply a binary system in contact in recent history, {\it and} considering the new radial velocity measurements,  
the likely conclusion is that DY\,Cen is not a spectroscopic binary  but is subject to pulsations affecting the profiles of photospheric lines. 

Close inspection of the UVES spectra shows that that strong stellar absorption line profiles vary over the course of the observations in a manner consistent with radial pulsations
\citep{montanes01}, {\it i.e.} the line cores shift red and blue more than the wings, creating triangular shaped profiles (Fig.~\ref{f:lpvs}).
These variations can account entirely for the changes in radial velocity measured in 2012. 
Similar variations are present but less obvious in the noisier 2015-16 observations. 
Profile variations are not evident in the C{\sc iv} lines discussed by \citet[][Fig.\,2]{rao12} and which are significantly weaker than those shown here in Fig.\,\ref{f:lpvs}.
Regarding other observations listed in Table\,\ref{t:vels}, only the UVES and HRS spectra have the S/N, resolution and number to show the form and variability of the line profiles. 

Irregular pulsations on a timescale of a few days have been observed in photometry of DY\,Cen \citep{pollacco91a} and also in other luminous helium stars such as PV\,Tel, FQ\,Aqr and V2205\,Oph \citep{walker81,jeffery85a,jeffery85b}. 
ASAS-SN photometery of DY\,Cen during 2016 -- 2018 continues to show irregular variability with an amplitude (2$\sigma$) of some 0.12 mag \citep{shappee14,jayasinghe19}. 
In the case of V2205\,Oph, a radial velocity range $> 10 \kmsec$ was measured from line core positions over an interval of 4\,n, 
accompanied by line profile variations indicative of non-radial pulsation \citep{jeffery92}.  
The present data are too sparse to establish whether the pulsations in DY\,Cen are radial or non-radial. 
Continued study of DY\,Cen's radial velocity and line profiles is desirable.


\begin{table*}
\begin{center}
\caption{Spectroscopic analysis of the evolving spectrum of DY\,Cen. 
Errors on final digits for \Teff,  $\log g$ and $v \sin i$ and mean abundances are given in parentheses. 
Abundances are given as $\log \epsilon$, normalised to $\log \Sigma \mu \epsilon = 12.15$.
Strontium abundances in square parentheses are estimates from spectral syntheses (see Fig.~\ref{f:lines}).  Errors in the last significant figure(s) are given in parentheses.  }  
\label{t:abunds}
\begin{tabular}{l rrrr  rrr }
\hline
Telescope & ESO 3.6m & AAT & VLT UT2 & SALT & &    Sun & DY\,Cen  \\
Instrument & CASPEC & UCLES & UVES  & HRS   &     & as09 & -- Sun  \\
Year  & 1987  & 2002  & 2010 & 2015 & mean   &   \\[1mm]
\teff /\,kK & 18.8(2) & 21.5(2) & 23.4(1) & 24.4(3) &  &    \\
$\sg / \cmss$  & 1.89(2) & 2.22(2) & 2.46(2) & 2.60(3) & &   \\[1mm]
C     & 9.56 & 9.19 & 9.22 & 9.39 & 9.34(17) & 8.43 & 0.89 \\
N     & 8.12 & 8.20 & 8.08 & 7.90 & 8.07(13) &  7.83 & 0.24 \\
O     & 9.05 & 9.04 & 9.00 & 9.09 & 9.05(04) &  8.69 & 0.36 \\
F     & ---  &  ---   & 6.70 & --- & 6.7(2) & 4.56 & 2.14 \\
Ne    & --- & 8.24 & 8.23 & 8.35 & 8.27(07) &  7.93 & 0.34 \\
Mg    & 7.95 & 7.60 & 7.94 & 7.18 & 7.66(36) &  7.60 & 0.06  \\
Al    & 6.43 & 6.13 & 5.89 & 5.76 & 6.05(29) &  6.45 & --0.40 \\
Si    & 7.10 & 7.14 & $^b$6.24 & 7.08 & 7.11(03) &  7.51 & --0.40 \\
P     & 5.87 & 5.82 & 5.77 & 6.11 & 5.89(15) &  5.41 & 0.48 \\
S     & 6.05 & 6.55 & 6.21 & 6.63 & 6.36(27) &  7.12 & --0.75 \\
Fe    & 6.84 & 6.48 & 6.57 & --- & 6.63(19) &  7.50 & --0.87 \\
Sr    & 6.98 & [6.5] & [6.5] & --- & 6.9(2) &  2.87 & 4.15 \\[1mm]
$v_{\rm rot} \sin i/ \kmsec$ & 20(5) & 30(5) & 30(5) & 40(5) &  \\
$v_{\rm crit}/ \kmsec$  & 80 & 97 & 111 & 120 & \\
\hline
\end{tabular}\\
Notes. as09: \citet{asplund09},
$b$: Si{\sc iii} lines not observed, excluded from mean.
\end{center}
\end{table*}

\begin{figure*}
\epsfig{file=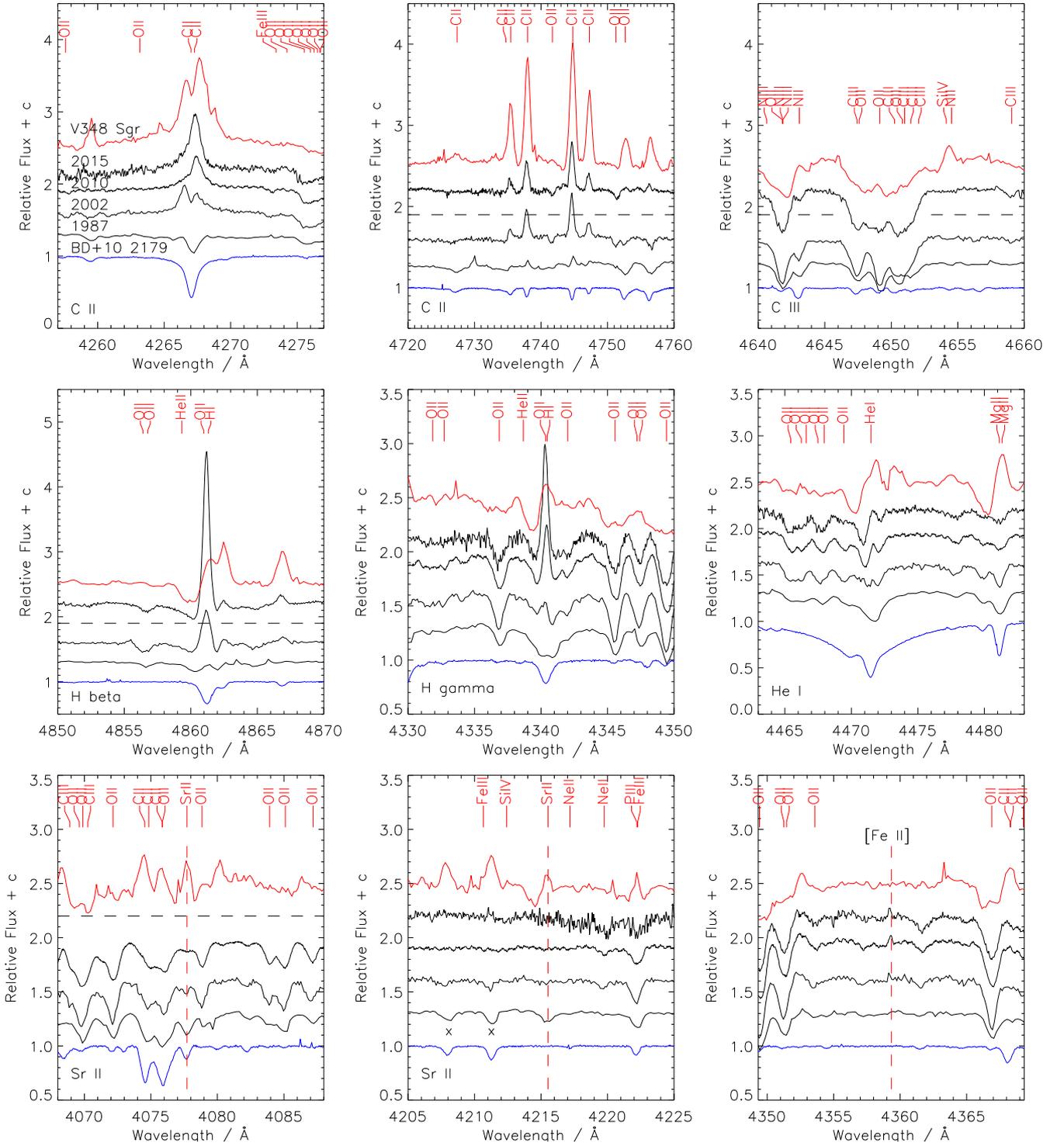,width=178mm,clip=,angle=0}
\caption{The evolving spectrum of DY\,Cen as indicated by representative C{\sc ii}, C{\sc iii},  H, He{\sc i}, Sr{\sc ii} and [Fe{\sc ii}] lines. 
Data are shown for  \'echelle spectra from 1987 (ESO/CASPEC), 2002 (AAT/UCLES), 2010 (VLT/UVES) and 2015 (SALT/HRS) (labelled: top left) panel). 
For comparison, spectra of the extreme helium star BD$+10^{\circ}2179$ (bottom / blue: ESO/FEROS) \citep{kupfer17} and the [WC12] planetary nebula central star V348\,Sgr (top / red: 1987 ESO/CASPEC) \citep{leuenhagen94} are also shown. 
Line identifications are for predicted lines from the best-fit model for the 2002 AAT/UCLES spectrum of DY\,Cen (Table \ref{t:abunds}). 
Not all spectra cover all windows; the continuum level of missing regions is represented by a horizontal dashed line. 
Lines marked '$\times$' are unidentified  in our own tables, by \citet{kupfer17} and in the NIST Atomic Spectra Database \citep{nist18}.
Broken vertical lines highlight selected features. 
} 
\label{f:spec}
\end{figure*}

\begin{figure*}
\epsfig{file=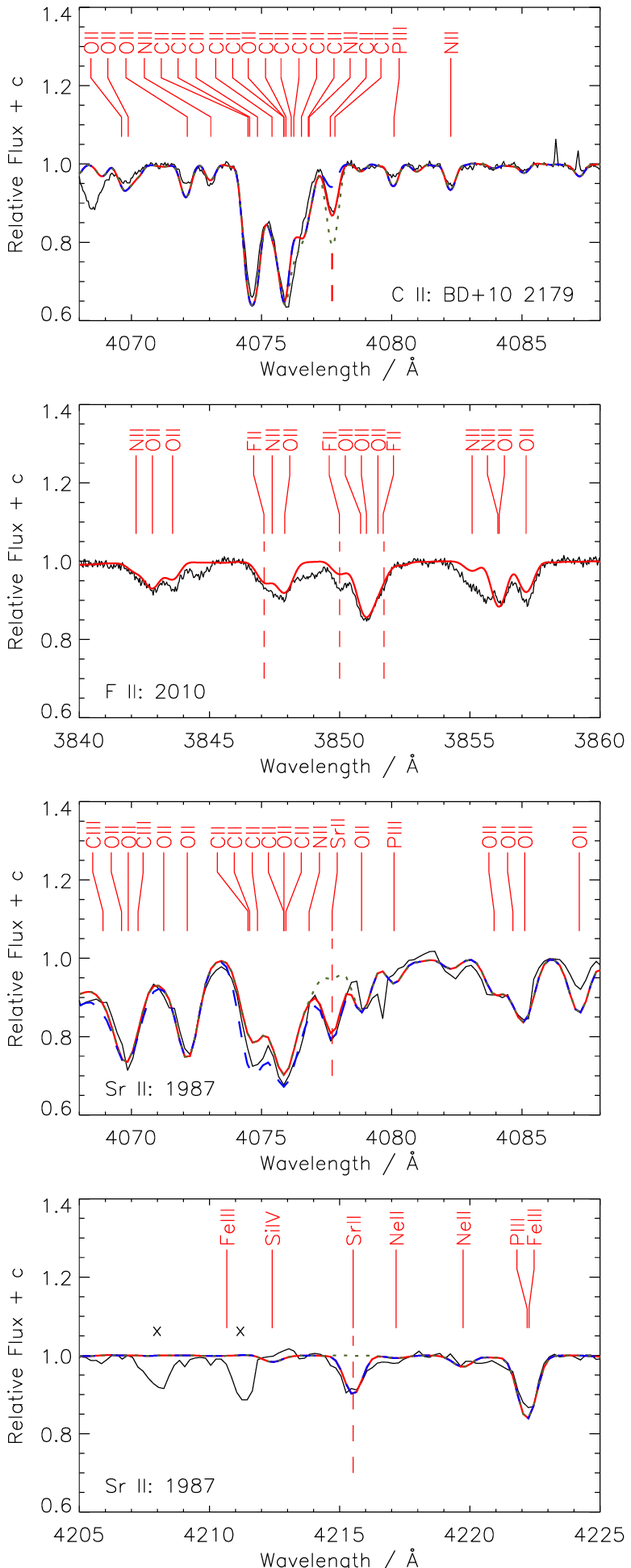,width=80mm,clip=,angle=0}
\epsfig{file=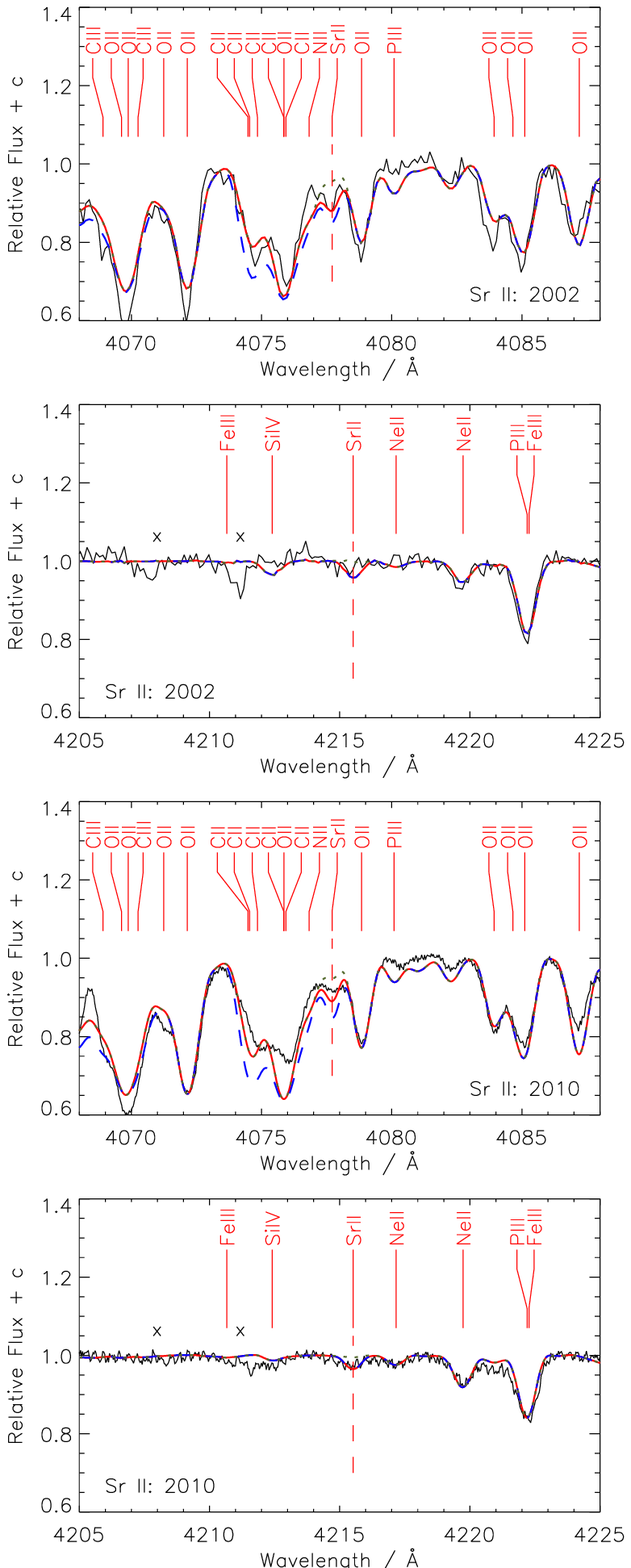,width=80mm,clip=,angle=0}
\caption{Fits to the spectra of DY\,Cen around  F{\sc ii} and  Sr{\sc ii} lines
for the  1987 (ESO/CASPEC), 2002 (AAT/UCLES) and 2010 (VLT/UVES)  \'echelle spectra, respectively.  
Lines marked '$\times$' are unidentified. The observed spectrum is a solid black line. 
The best-fit model with the abundances shown in Table 2 is solid red. For 1987, the same model with the 
carbon abundance {\it enhanced} by 0.5 dex is dashed blue. With strontium {\it reduced} by 1 dex it is dotted green. 
The model strontium abundances are, by year, $\log \epsilon_{\rm Sr} = 6.92 (1987), 6.5 (2002)$ and 6.5 (2010). 
The top left panel shows the spectrum of BD$+10^{\circ}2179$ (black) in the vicinity of Sr{\sc ii} 4077\AA\ (vertical dashed line).  Model spectra computed for the abundances and other parameters obtained by \citet{kupfer17} are shown with the C{\sc ii} $gf$ values given by \citet{kurucz_cd23} (green dots), \citet{K14} (blue dashes) and as adopted here (red). There is no contribution from strontium.  } 
\label{f:lines}
\end{figure*}

\begin{figure}
\epsfig{file=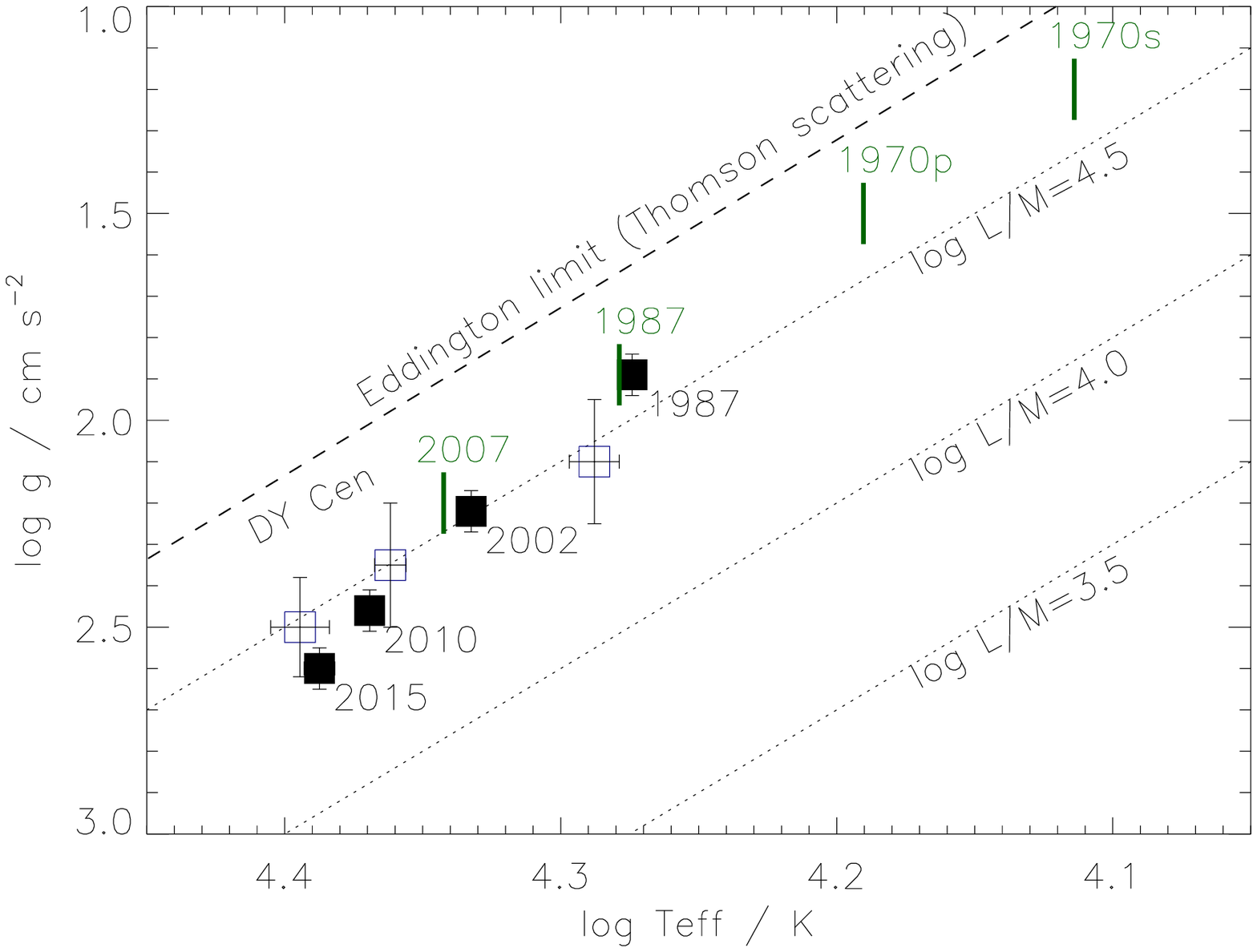,width=80mm,clip=,angle=0}
\caption{The evolution of  DY\,Cen as a function of \Teff\ and \sg\ as measured from \'echelle spectroscopy (open navy squares) and photometry (green vertical bars and labels) by \citet{pandey14} and in the current paper (solid black squares, labelled). 
Two colour temperatures for 1970 are from \citet{pandey14} (1970p) and \citet{schaefer16} (1970s). Diagonal  lines represent the Eddington limit for Thomson scattering (dashed) and contours in $L/M$ (dotted). }  
\label{f:gt}
\end{figure}

\section{Spectroscopic Analysis}
\label{s:spec}
The SALT/HRS spectrum provides an additional reference point at which to monitor the contraction of DY\,Cen. 
We used a spectrum synthesis approach based on LTE plane-parallel model atmospheres 
\citep{behara06}, to measure the effective temperature, surface gravity, and surface composition from all four epochs of high-resolution \'echelle spectra available to us.
A grid of model atmospheres and emergent spectra was computed (in local thermodynamic equilibrium: LTE) assuming a composition approximately as given by \citet{pandey14} and a microturbulent velocity of 20\,\kmsec. 
A best fit to the coadded spectrum from each epoch was found by interpolation,  yielding the effective temperature (\teff) and surface gravity ($g$) shown in Table\,\ref{t:abunds}.
The grid model most closely matching these values was used as input to a second optimization step in which the abundances of observable species were obtained  by minimizing the difference between theoretical and observed spectra \citep{jeffery01b}.
Abundance errors are estimated using 
the standard deviation around the mean of measurements made at all four epochs (where possible) on the basis that the surface composition has not changed between 1987 and the present.

Since the hydrogen abundance is extremely difficult to obtain at later epochs owing to increasing emission, the number fractions of hydrogen and helium were fixed at 10\% and 88\% as measured by \citet{jeffery93a}. 
The projected rotation velocity ($v_{\rm rot} \sin i$) is an additional parameter of both optimization procedures and is determined from the profiles of all lines simultaneously. 
Overall results are shown in Table 2, together with the mean abundance, the model abundance, previous measurements by \citet{jeffery93a} and \citet{pandey14}, and the solar value \citep{asplund09}.
Errors reported in Table 2 are all formal errors. 
Formal errors on individual abundance measurements are $\approx 0.05$. 
Continuing evolution towards higher \teff\ at a rate of approximately 100\,K yr$^{-1}$ is demonstrated by the most recent data  (Fig. \ref{f:gt}).  

There are systematic differences in \teff\ and $\log g$ between the present results and \citet{pandey14} (Fig. \ref{f:gt}), with the former being consistently cooler than the latter. The principal reasons are the former's use of partially-blanketed non-LTE models and a line-by-line approach, compared with the current use of fully-blanketed LTE models and a spectrum-synthesis approach. 
Both have merits, and both are subject to the available spectral range and choice of absorption lines. 

Crucially, the relative shifts between epochs are similar and, to within errors, both sets of results have similar luminosity-to-mass ($L/M$) ratios. 
The difference which is harder to account for is that $L/M$ remains roughly constant in \citet{pandey14}, whereas we find it to fall steadily by $\sim 0.25$ dex in 30 years. 
The primary suspect is the manner in which surface gravity is measured principally from Stark-broadened He{\sc i} lines; these are increasingly affected by emission (Fig.\,\ref{f:spec}). 

\begin{figure}
\epsfig{file=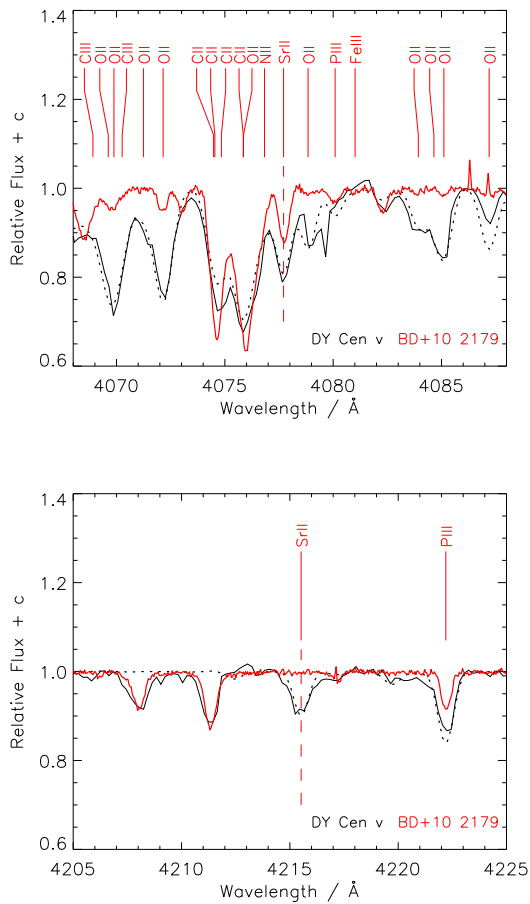,width=80mm,clip=,angle=0}
\caption{The spectrum of DY Cen  in 1987 (black, solid) in the region of the Sr{\sc ii} lines at 4077.7 and 4215.5\AA\, compared with the spectrum BD$+10^{\circ}2179$ (red). The best-fit model for DY\,Cen (1987) is also shown (dotted), and the most significant lines contributing to the model are labelled.}  
\label{f:sr}
\end{figure}

\subsection{Strontium}
An advantage of the spectrum-synthesis approach is that it reveals absorption lines that are either absent from the linelist or have been incorrectly assumed to have a low abundance. 
In our spectra, several features are either due to instrumental or data reduction artefacts, but several are due to absorption lines that have not yet been identified. 
In the 1987 CASPEC spectrum, two of these turn out to be the singly-ionized strontium lines at 4077.7 and 4215.5 \AA\ (Fig. \ref{f:spec}).
These were unexpected and overlooked by \citet{jeffery93a}. 

The Sr{\sc ii} 4077.7\AA\ line is blended with a group of C{\sc ii} lines \citep[$3d\,^4D^{\circ} - 4f'\,^4F$:][]{K14}.
The strongest of these are seen in both the spectrum of DY\,Cen and of an EHe with similar \Teff, BD$+10^{\circ}2179$ \citep{kupfer17} (Fig.\,\ref{f:sr}).
There is no evidence for Sr{\sc ii} 4215.5 \AA\ in  BD$+10^{\circ}2179$ (Fig.~\ref{f:spec}),
and we conclude that Sr{\sc ii} 4077.7 \AA\ does not contribute to the C{\sc ii} blend in this star. 
All C{\sc ii} lines belonging to the multiplet are included in the spectrum syntheses.  
Atomic data for these lines were checked against the Vienna Atomic Line Database \citep{vald95,vald15}. 
For two lines, namely C{\sc ii} 4077.641 and 4077.824 \AA, there is a discrepancy $>1$ dex between $gf$ values given by \citet{kurucz_cd23} and \citet{K14}. 
Assuming $gf$ values for all other lines in the multiplet are approximately correct, both sets of theoretical $gf$ values for the 4077 \AA\ lines are inconsistent with the line ratios observed in BD$+10^{\circ}2179$ (Fig.~\ref{f:lines}). 
To achieve consistency, we adopted empirical $gf$ values as in Table~\ref{t:gf} for the discrepant lines and from \citet{K14} for other multiplet members.

\begin{table}
    \centering
    \begin{tabular}{ccccc}
\hline
Ion & $\lambda/$\AA\ &  & $\log gf$ & \\
    &          & KB & K14 &  Empirical  \\
    \hline
C{\sc ii} & 4077.641 & $-0.500$ & $-1.511$ & $-0.950\pm0.05$ \\
          & 4077.824 & $-0.200$ & $-1.841$ & $-0.650\pm0.05$ \\
          \hline
    \end{tabular}
    \caption{Published and empirical $gf$ values used in this analysis. KB: \citet{kurucz_cd23}, K14: \citet{K14} }
    \label{t:gf}
\end{table}

There is a very weak C{\sc ii} line predicted at 4216.55\AA\ \citep[$5d\,^2D - 3d'\,^2P^{\circ}$:][]{K14}. 
It is not seen in either DY\,Cen, BD$+10^{\circ}2179$, or in any of the synthetic spectra we have computed.   
To be visible in a mid B-type spectrum with significant rotational broadening, the strontium abundance must be very large. 
From the 1987 CASPEC spectrum, we measure $\log \epsilon_{\rm Sr} = 6.92\pm0.15$, more than 4 dex above solar (Fig.\,\ref{f:lines}). 
Both Sr{\sc ii} lines are present in the 2002 AAT and 2010 UVES spectra, but are either blended or weak. 
Satisfactory comparisons between theory and observations are only obtained with $\log \epsilon_{\rm Sr} \geq 6.5$ in both cases. 
The decrease in the strength of both lines is completely consistent with the increasing temperature, as demonstrated in Fig.\,\ref{f:lines}. 
To be confident a) of the very high strontium abundance and b) that the observed line is not being contaminated by nearby carbon lines, Fig.\,\ref{f:lines} shows models computed with (a) the strontium abundance {\it reduced} by 1 dex and (b) the carbon abundance {\it increased} by 0.5 dex.  
The observed lines in all three years from 1987 to 2010 can only be accounted for by a strontium abundance $\approx4$ dex above solar. 

There is no indication of other s-process elements, notably yttrium or zirconium, in the spectrum. 
Upper limits may be set for yttrium at $<3.1$ dex above solar and for  zirconium at $<1.9$ dex above solar by the non-detection of Y{\sc iii} 4039.6\AA\ and Zr{\sc iv} 4198\AA\ respectively. These are the strongest yttrium and zirconium lines in the spectral range and at the effective temperature of DY\,Cen over the period of the high-resolution observations.  
The limits are obtained by examination of the 2010 spectrum where both lines should be stronger than in the earlier observations and by requiring a detectable line to have equivalent width $>15$ m\AA, corresponding to the weakest identified lines in the 2010 spectrum. 

\subsection{Emission}
The hydrogen, helium and carbon emission spectra continue to strengthen  substantially over time (Fig.\,\ref{f:spec}).
We note that the Balmer series is visible in emission in the VLT/UVES spectrum  up to $n=12$ (3683\AA), and that by 2010 many C\,{\sc ii} lines are increasingly affected by emission. 

\section{Discussion} 

The availability of high-quality spectra covering nearly three decades allows us to see,  in real time, the transformation of a star from one resembling an extreme helium star dominated by photospheric absorption lines (i.e. BD$+10^{\circ}2179$) with spectral type early B(He), to one resembling a cool Wolf-Rayet central star of a planetary nebula (i.e. V348\,Sgr) with spectral type [WC12] (Fig.\,\ref{f:spec}). This dynamic appears to be reflected by changes in the surrounding nebula, with evidence that it formed about 1100\,yr ago \citep{rao13}. 

DY\,Cen is not a binary, it is rich in strontium, it is evolving fast, and its surface rotation is increasing.

The surface  of DY\,Cen is C and O enriched, and depleted in iron and most other light elements, and has been discussed extensively by \citet{jeffery93a} and \citet{pandey14}. 
The strong enhancement of fluorine points to a connection with extreme helium stars (EHes) \citep{pandey06} and R\,CrB (RCB) variables \citep{pandey08}. 
\citet{jeffery11a} argued the fluorine excess in EHes to be an intershell product from prior evolution on the asymptotic giant branch (AGB), exposed on the surface by  mixing during the merger of two white dwarfs.
\citet{pandey14} argued the excess in DY\,Cen to be evidence of a short-exposure of \iso{18}{O} to hot protons during the merger process.    
Either would contradict evidence for a 40\,d orbital period, but are more easily reconciled with DY\,Cen being a single star. 
However, DY\,Cen is substantially more hydrogen rich than any other EHe or cool RCB.  

The strontium excess points to a period of s-process enrichment either whilst on the asymptotic giant branch (AGB) or during a single strong neutron-exposure event. 
DY\,Cen is not currently a post-AGB star; its evolution is too rapid and the surface composition too hydrogen-poor. 
The high strontium abundance could point to an explanation for the apparent low metallicity. If neutron captures on iron account for the very high strontium abundance, which is comparable with iron, then the original iron abundance would have been substantially higher. 
Otherwise strontium poses a puzzle.

EHes and RCBs are now commonly considered post-merger double white dwarfs \citep{dan14,staff18,menon19}.
The question arises whether these are related to the handful of stars that includes DY\,Cen, V348\,Sgr, MV\,Sgr and HV\,2761  and which share the property of now being hot and of being or having been RCB variables \citep{demarco02,schaefer16}.
The alternative is that they are the contracting remnants of post-AGB stars which re-ignited helium whilst approaching the white dwarf cooling sequence, such as FG\,Sge and V4334\,Sgr  \citep{schoenberner79,asplund00,herwig01,jeffery06}. 
Together with the yttrium and zirconium upper limits, the strontium abundance is not inconsistent with the Sr/Y/Zr abundance pattern of V4334\,Sgr (5.4/4.2/3.5 dex) \citep{asplund97b}.

As Schaefer noted, DY\,Cen has evolved from being a cool RCB star in around 1907 to resemble an EHe in 2014, its high surface hydrogen notwithstanding.  
DY\,Cen (and possibly V854\,Cen) is also unusual amongst RCB stars by showing signatures of polyaromatic hydrocarbon and fullerene molecules, either produced by high-temperature condensation or from decomposition of hydrogenated amorphous carbon ejected when the star was considerably cooler \citep{garcia11a}. 

The heating rate of DY\,Cen measured here and by \citet{pandey14} is $\approx 200 {\rm \,K\,yr^{-1}}$. 
The heating rates indicated for EHes V2076\,Oph, PV\,Tel, V2205\,Oph and NO\,Ser are $\approx 120, 20, 95$ and $33 {\rm \,K\,yr^{-1}}$ \citep{jeffery01c}, the higher values being only a factor 2 smaller than DY\,Cen. 

The heating rate for a star with a degenerate core depends on the thermal timescale of the envelope  ($\propto M_{\rm env}/L$).
For post-shell-burning stars, the luminosity can be related to the core mass ($M_{\rm c}$) by $L \propto M_{\rm c}^{\delta}$, $\delta$ varies from 2.4 ($M_{\rm c}=0.8\Msolar$) to 1.65 (1.0\Msolar)  \citep{jeffery88b}.
To reduce the contraction timescale by a factor 2 requires $M_{\rm c}$ to increase by
 $\times \sqrt 2$ or, more likely, to reduce $M_{\rm env}$ by a factor 2.  

The Gaia Data Release 2 parallax for DY\,Cen is $0.0500\pm0.0230$ mas \citep{gaia18.dr2}.
This is on the margin of where Gaia distances are currently reliable and corresponds to a distance of some $11.9+2.0/-2.6$\,kpc.  
While Martin \& Jeffery (in prep.) present an analysis of distances and kinematics of EHes and related objects, including DY\,Cen, their accuracy is not yet sufficient to constrain the mass of DY\,Cen and hence establish whether its contraction is due to high $M_{\rm c}$ or low $M_{\rm env}$. 

A significant feature in the spectrum of DY\,Cen is the increasing width of isolated spectral lines with time (e.g. O{\sc ii} 4367: Fig.\ref{f:spec}). 
This exceeds the increase in thermal broadening and translates to an increase in projected rotation velocity \vsini. 
 
As DY\,Cen heats, its surface layers contract and, to conserve angular momentum, they must spin up, as observed.
Since the envelope is radiative, surface layers will rotate differentially, with maximum rotation at the equator.
Table\,\ref{t:abunds} compares \vsini\ with the critical rotation velocity $v_{\rm crit}$ for a star of mass $M=0.9\Msolar$, and demonstrates that the ratio $v_{\rm rot} /v_{\rm crit}$ is increasing and may already be approaching unity if $\sin i < 0.5$.
Increasing rotation reduces the effective equatorial surface gravity which in turn will enhance the equatorial wind, as appears to be demonstrated by the strengthening emission in H, C{\sc ii} and He{\sc i} (Fig.\ref{f:spec}). 
If the star cannot shed sufficient mass by an equatorial wind, more violent mass-loss must ensue as it reaches breakup. 
Rotational breakup for contracting EHes was anticipated for post white dwarf merger models by \citet{gourgouliatos06}. 
On the grounds of angular momentum conservation, the high rotation rate in DY\,Cen seems unlikely to be a consequence of late thermal pulse evolution if the star was significantly smaller than it is currently at the time of the hydrogen-shell flash. However, that apparent contradiction may not arise if a) the thermal pulse occurred before the post-AGB progenitor had contracted onto the white dwarf cooling track, b) the star lost mass at maximum radius and thus reducing the critical rotation rate for subsequent evolution, or c) additional angular momentum was transferred into the envelope from a more rapidly rotating degenerate core by convective mixing following the thermal pulse. 

Continued contraction will lead the spectrum to further resemble that of V348\,Sgr; it is possible that RCB-type activity could recommence. 
Evidence for this will come from close monitoring of photometric behaviour. 
Unless there is a mode change, the pulsation period ($\Pi$) should decrease along with the radius ($R$)  since $\Pi\propto1/\sqrt{\bar{\rho}}\propto R^{3/2}\propto\teff^{-3}$ at constant $L$.
With $\Pi = 3.8 - 5.5$d in 1987 \citep{pollacco91a}, it should now be less than half this value.  
Pulsation and rapid rotation may precipitate minor dust-formation episodes as precursors to full RCB-type minima.

Within approximately 30 years, \Teff\ will exceed 30\,000\,K. The increasing far-UV flux will further ionize gas surrounding the star, perhaps exciting a highly structured planetary nebula and providing new information on the history of DY\,Cen. 

\section{Conclusion}
\label{s:conc}
 
New time-resolved spectroscopy of DY\,Cen covering several period ranges demonstrates no evidence for it being a spectroscopic binary. 
Our previous claim for a 39.7\,d period \citep{rao12} was probably consequent on an unfortuitous distribution of historical observations obtained over several decades,  confusion with surface motion due to pulsations, and undiagnosed systematic errors. 

Self-consistent analysis of high-resolution spectroscopy covering four epochs from 1987 to 2015 demonstrates continuing rapid blueward evolution, corresponding to contraction at near-constant luminosity.  
Whilst, with the exception of hydrogen, the surface composition is similar to that of EHes and RCBs, the rate of blueward evolution is a factor of two higher. 
Contraction is associated with rotational spin-up approaching 50\% of the critical rate. 

The unexpected discovery of high levels of strontium in the 1987 spectrum provides exciting new data for testing evolution models for DY\,Cen. Whilst the s-process  certainly contributes to the surface chemistry of EHes and RCBs \citep{lambert94,jeffery11a}, it also plays a r\^ole in the surface chemistry of late thermal pulse stars such as FG\,Sge and V4334\,Sgr  \citep{jeffery06,asplund00}. 

DY\,Cen and objects like it provide a remarkable astrophysical laboratory in which to measure abundances of different elements simply as a consequence of its changing effective temperature. 
It will be important to continue to obtain high-quality spectroscopy of DY\,Cen at all wavelengths as its surface continues to heat.

\section*{Acknowledgments}
This paper includes data collected at the Southern Africa Large Telescope, the European Southern Observatory and the Anglo-Australian Telescope. 
They authors thank Dr Steven Crawford, formerly of the SAAO, for painstaking help with reduction of SALT/HRS data.  

The Armagh Observatory and Planetarium is funded by direct grant form the Northern Ireland Dept for Communities.
The authors acknowledge support from the UK Science and Technology Facilities Council (STFC) Grant No. ST/M000834/1. 

This work has made use of the VALD database, operated at Uppsala University, the Institute of Astronomy RAS in Moscow, and the University of Vienna.
\bibliographystyle{mn2e}
\bibliography{ehe}

\appendix
\renewcommand\thefigure{A.\arabic{figure}} 
\renewcommand\thetable{A.\arabic{table}} 

\section[]{Spectral Atlas  for DY\,Cen}
\label{s:app1}
\label{s:lines}
\begin{figure*}
\epsfig{file=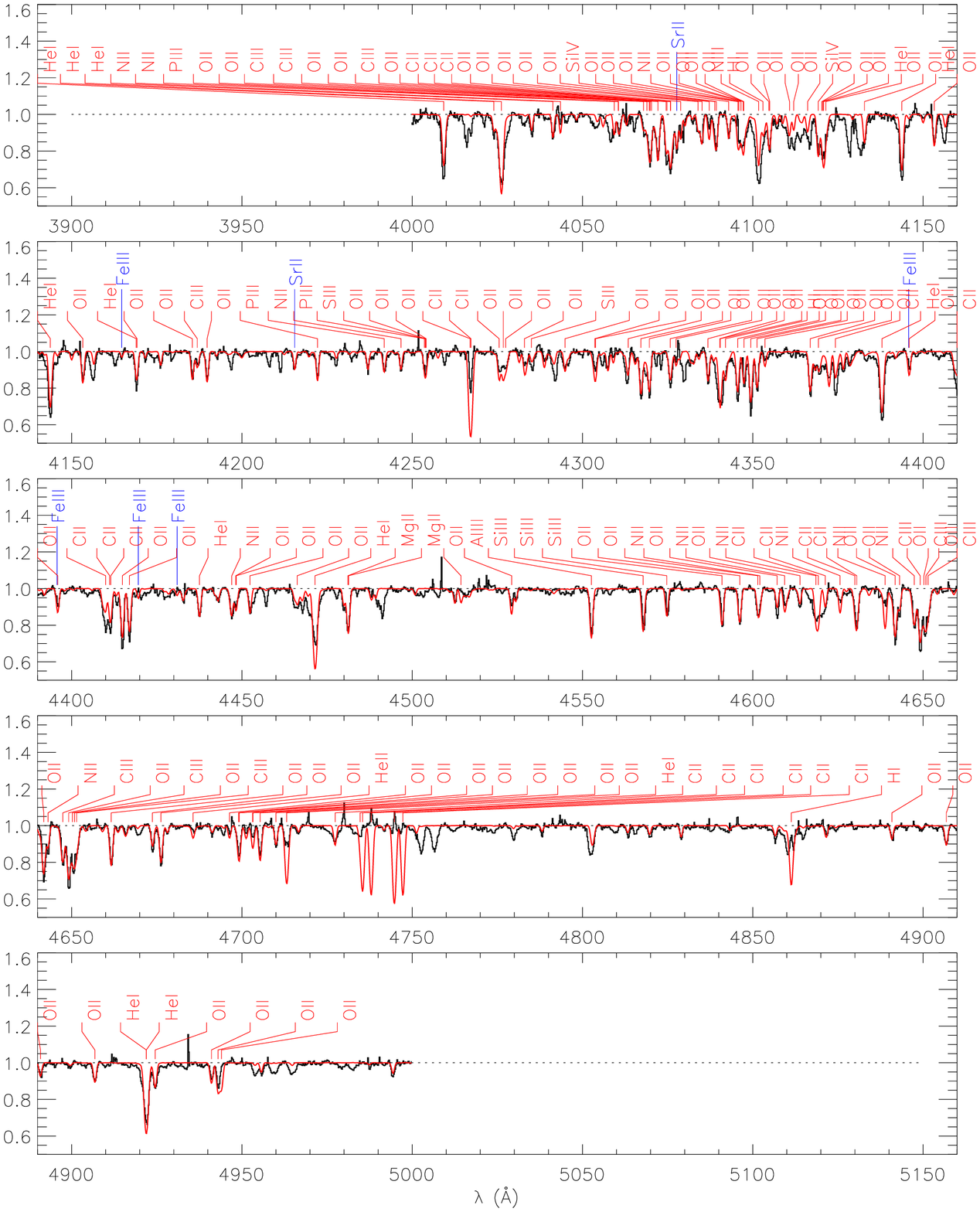,width=0.9\textwidth}
\caption{1987 ESO 3.6m/CASPEC spectrum of DY\,Cen (black histogram), 
and the best-fit model having  $\teff=19\,000$\,K, $\lgcs=1.9$,  $n_{\rm H}/n_{\rm He}=0.11$ 
and abundances shown in Table~\ref{t:abunds} (red polyline).
Lines with theoretical equivalent widths $W_{\lambda}7$m\AA\ are labelled. }
\label{f:atlasA}
\end{figure*}

\begin{figure*}
\epsfig{file=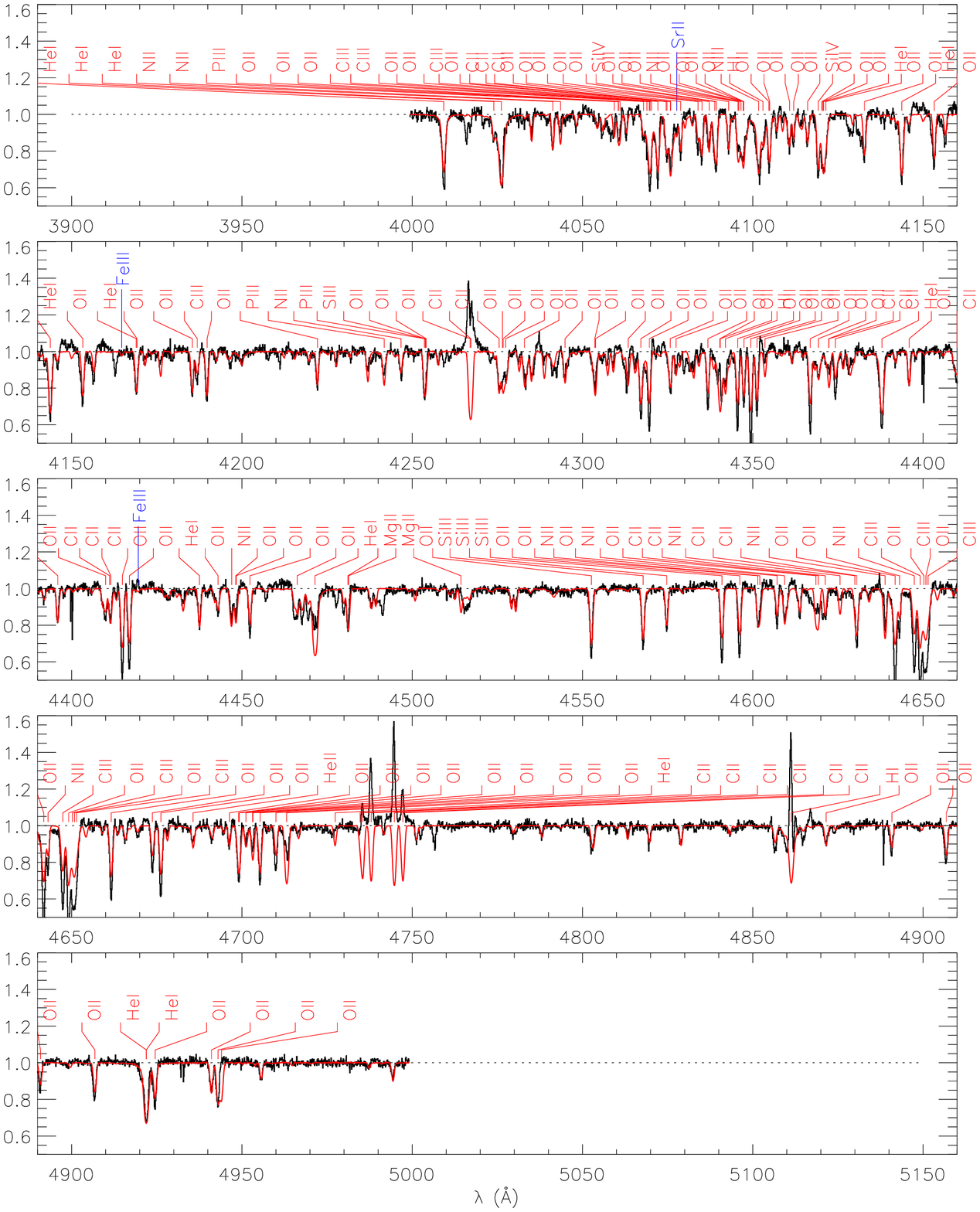,width=0.9\textwidth}
\caption{As Fig.\,\ref{f:atlasA} for the 2002 AAT/UCLES spectrum, $\teff=21\,500$\,K, $\lgcs=2.20$ and lines with $W_{\lambda}>10$m\AA. }
\label{f:atlasB}
\end{figure*}

\begin{figure*}
\epsfig{file=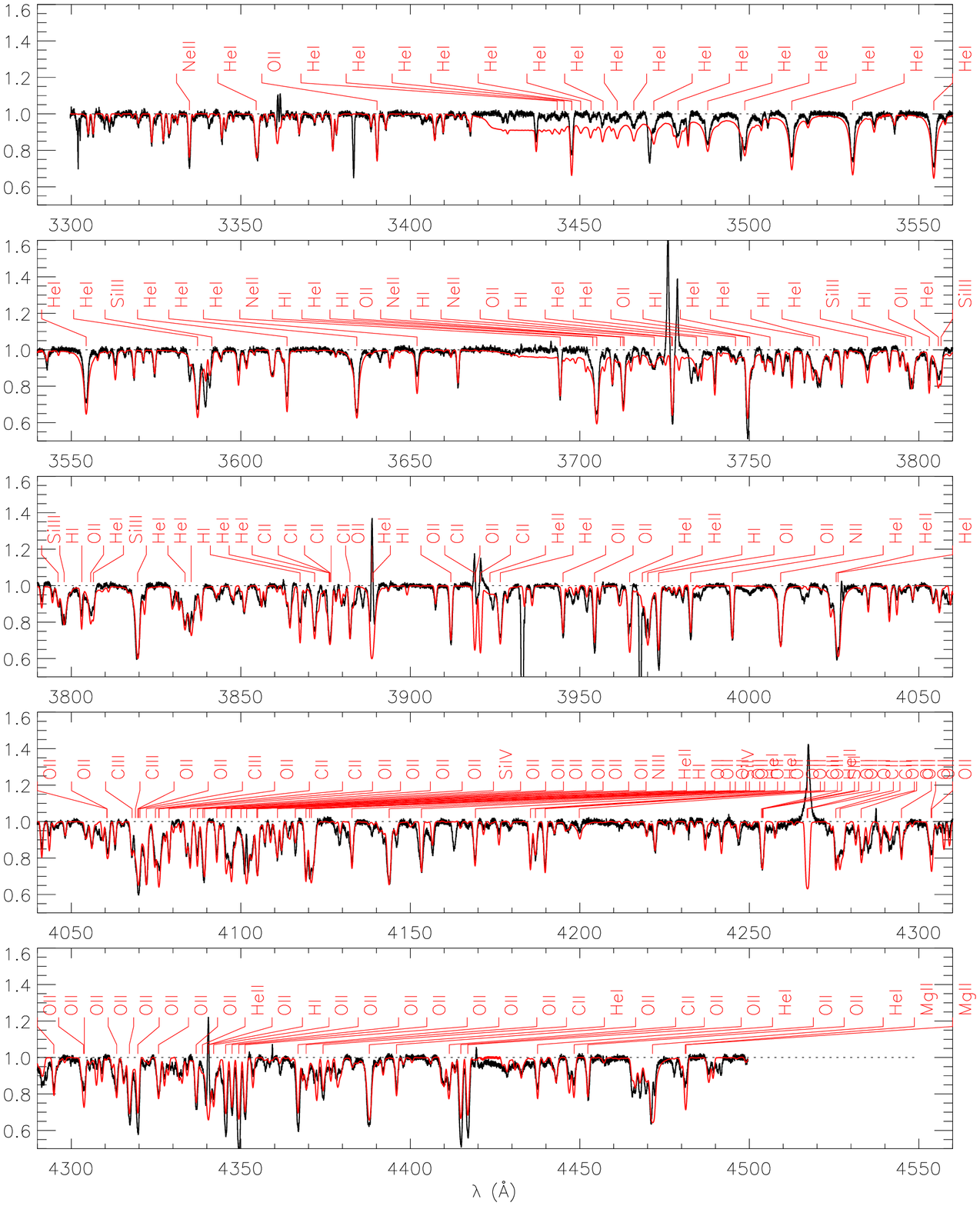,width=0.9\textwidth}
\caption{As Fig.\,\ref{f:atlasA} for the coadded 2010 VLT/UVES spectrum,   $\teff=23\,500$\,K, $\lgcs=2.50$, and lines with $W_{\lambda}>15$m\AA. 
Mismatches between observation and model around 3420 and 3670 A are due to the inclusion of high-order members of He{\sc i} series. The current
version of our spectrum synthesis code does not treat dissolution into the continuum and hence introduces a discontinuity at series limits when
these lines are included. 
}
\label{f:atlasC}
\end{figure*}

\begin{figure*}
\epsfig{file=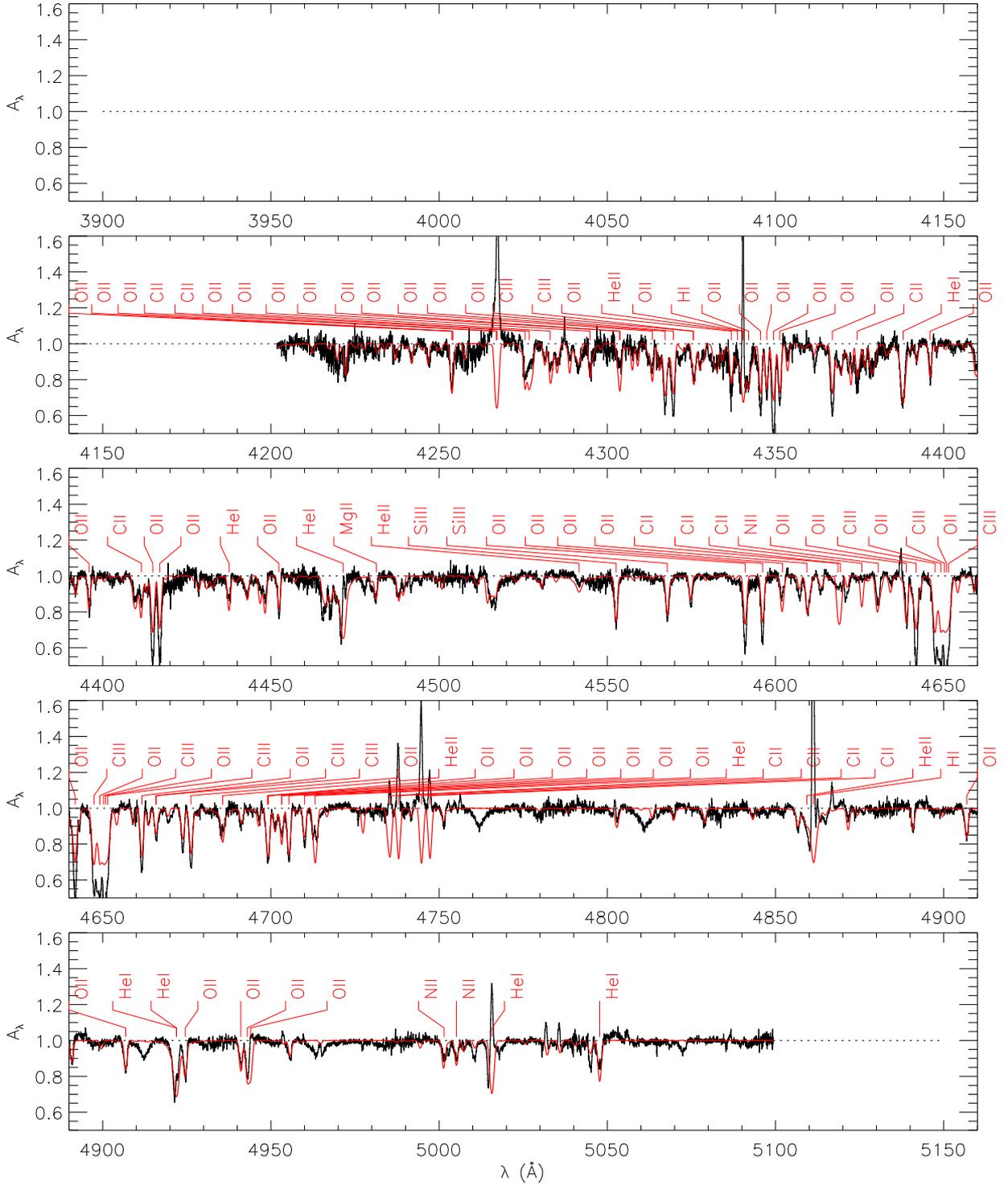,width=0.9\textwidth}
\caption{As Fig.\,\ref{f:atlasA} for the coadded 2015/16 SALT/HRS spectrum, $\teff=24\,500$\,K, $\lgcs=2.60$, and lines with $W_{\lambda}>15$m\AA. The broad features around 4760, 4810 4910 and 4965 \AA\ are artefacts of the data reduction process.  }
\label{f:atlasD}
\end{figure*}

Figures \ref{f:atlasA} to \ref{f:atlasD} contain an  atlas of the \'echelle spectra of DY\,Cen obtained in 1987, 2002, 2010 and 2015/6 (and summed by epoch), together  with the best model fit and identifications of major absorption lines.  

\label{lastpage}
\end{document}